# On-chip Non-Hermitian Cavity Quantum Electrodynamics

Yan Chen[1,2], Xudong Wang[3], Jin Li[4], Rongbin Su[5], Kaili Xiong[2], Xueshi Li[2], Ying Yu[6,7], Tao Zhang[2], Kexun Wu[4], Xiao Li[1], Jiawei Wang[4]*, Jiaxiang Zhang[3]*, Jin Liu[6,8]*, Tian Jiang[1,2,9]*

[1]College of Advanced Interdisciplinary Studies, National University of Defense Technology, Changsha, 410073, China.
[2]Institute for Quantum Science and Technology, National University of Defense Technology, Changsha, 410073, China.
[3]Institute of Microsystem and Information Technology,
Chinese Academy of Sciences, Shanghai 200050, China.
[4]School of Integrated Circuits, Harbin Institute of Technology (Shenzhen), Shenzhen 518055, China.
[5]School of Integrated Circuits and Optoelectronic Chips,
Shenzhen Technology University, Shenzhen 518118, China.
[6]State Key Laboratory of Optoelectronic Materials and Technologies, Sun Yat-sen University, Guangzhou, 510275, China.
[7]Hefei National Laboratory, Hefei 230088, China
[8]Quantum Science Center of Guangdong-Hong Kong-Macao Greater Bay Area, Shenzhen, China.
[9]Hunan Research Center of the Basic Discipline for Physical States, Changsha, 410073, China

*Corresponding author. Email: wangjw7@hit.edu.cn (J.W.); jiaxiang.zhang@mail.sim.ac.cn (J.Z.); liujin23@mail.sysu.edu.cn (J.L.) and tjiang@nudt.edu.cn (T.J.)

**Abstract:** Exceptional points (EPs) promise revolutionary control over quantum light-matter interactions. Here, we experimentally demonstrate flexible and reversible engineering of quantum vacuum fluctuation in an integrated microcavity supporting chiral Eps. We develop a hybrid lithium niobate (LN)-GaAs quantum photonic platform, seamlessly combining high-quality quantum emitters, a low-loss photonic circuit, efficient electro-optic (EO) effect, and local strain actuator in a single device. Chiral EPs are implemented by dynamically tuning the coupling between the modes associated with a micro-ring resonator, resulting in anomalous spontaneous emission dynamic with a 7-fold modulation of the lifetime (120 ps to 850 ps). Meanwhile, we reshape single-photon spectra via cavity local density of states (LDOS) engineering and generate non-Lorentzian spectral profiles: squared-Lorentzian, Fano-like, and EP-induced transparency (EPIT), a suppression of emission at zero detuning. This work unveils exotic cavity quantum electrodynamics (cQED) effects unique to EPs and establishes a universal paradigm for non-Hermitian quantum photonics.

**Main Text:**

EPs, singular degeneracies unique to non-Hermitian systems, occur when two or more eigenvalues and their corresponding eigenstates coalesce, creating unconventional spectral topology and dynamical phenomena (*1,2*). Photonic systems have emerged as a versatile platform for engineering non-Hermitian physics due to their precisely tunable gain-loss profiles, enabling systematic exploration of EP-induced effects (*3-5*). In single micro/nano-resonators operating at an EP, eigenstate coalescence imposes strict chiral mode locking, resulting in unidirectional photon propagation with complete suppression of backward scattering (*6-8*). Recent advances in classical photonic systems have unveiled remarkable EP governed dynamics, including the photonic Aharonov-Bohm effect (*9-11*), asymmetric backscattering suppression (*12-14*), chiral perfect absorption (*6,15,16*), and magnetic-free optical nonreciprocity (*13,17*). Moreover, the integration of EP physics into photonic device architectures has demonstrated transformative potential, particularly in developing ultrahigh-sensitivity sensors (*8,18-20*) and topological on-chip lasers (*21-23*).

The implications of EPs extend profoundly into quantum optics, where second-quantized light-matter interactions acquire new dimensions under non-Hermitian conditions. The presence of EPs fundamentally modifies quantum interference patterns and photon statistics, introducing non-classical features unattainable in Hermitian systems (*24,25*). A frontier challenge lies in investigating quantum light-matter interactions at Eps - a regime where cQED undergoes radical reformulation (*26,27*). Theoretical breakthroughs predict that quantum emitters coupled to EP-engineered cavities exhibit extraordinary spontaneous emission enhancements, deviating remarkably from conventional Purcell effects (*28-30*). This anomalous behavior stems from the emergence of non-Lorentzian LDOS profiles at EPs, enabling unconventional photon-emitter coupling at the single-quantum level. Despite these compelling theoretical frameworks, experimental realization of quantum light-matter interaction at EPs remains elusive.

Leveraging the advances in the hybrid integrated quantum photonics and thin film LN(TFLN) photonics (*31-35*), we here demonstrate the first experimental non-Hermitian cQED through the chiral EP microcavity integrated with semiconductor QDs in a hybrid lithium niobate LN-GaAs photonic circuit. The LDOS of the cavity is controlled by a tunable phase shifter in the bus waveguide terminated by a mirror. The phase shifter is enabled by the cryogenic Pockels effect of LN (*36*). The QDs are brought into resonance with cavity modes via the reversed piezoelectric effect of LN (*37*). Unlike conventional Lorentzian emission, distinct single-photon spectra, including squared-Lorentzian line shape and Fano-like line shape, are observed in the non-Hermitian system. Notably, an intriguing phenomenon EPIT occurs at specific parameters: the spectral density vanishes at zero QD-cavity detuning, and the QD's emission almost vanishes entirely. Additionally, the EPs significantly modify the spontaneous emission process of the QD, with the lifetime FP dynamically and reversibly tuned from 120 ps to 850 ps. These results are in excellent agreement with theoretical analysis. Compared with Hermitian cavities, our work provides greater degrees of freedom to control the light-matter interaction and we envision that the versatile platform could offer unprecedented opportunities for non-Hermitian quantum photonics.

**Theoretical framework**

The demonstration is based on a hetero-integrated quantum photonic chip which is artistically sketched in Fig. 1a. It consists of a LN micro-ring resonator evanescently coupled to a waveguide terminated by a reflector at one port. An EO phase shifter is inserted between. A tapered GaAs waveguide with embedded QD is transfer-printed on the LN resonator, forming a hybrid photonic circuit. The QD is a two-level quantum system and constitutes a cQED system together with the LN ring resonator. The QD's emission wavelength is strain-tuned via the piezoelectric effect to align with the cavity resonance. Details about the QDs' fabrication is described in Supplementary Materials section 1.

In a conventional micro-ring cavity (upper left panel of Fig. 1b), optical excitation of the QD by an external laser pulse generates a photon that coherently couples to two independent cavity modes i.e., the CW and CCW cavity modes and then dissipates to the evanescently coupled WG. Specifically, at the diabolic point (DP) when these two modes show degeneracy, the LDOS exhibits Lorentzian-shaped spectra.

For an EP microcavity (lower left panel of Fig. 1b), a deterministic unidirectional coupling term κ (from the CW to CCW mode in our case) is introduced by the on-chip reflector. The coupling coefficient can be derived by $\kappa = -2i\gamma_\kappa |r|e^{i\varphi}$, where $\gamma_\kappa$ is the coupling loss to the bus Waveguide; |r| is the field reflection amplitude, and φ is the phase shift induced by the phase shifter. The reflector is assumed to cause no phase change; otherwise, such a phase change can be included in φ. Upon one pumping cycle, the excitation can be stored in either the CW mode, CCW mode, or in the QD, which we denote as |1, 0, g⟩, |0, 1, g⟩, and |0, 0, e⟩ states. Using these three states as projection basis, the system's Hamiltonian is expressed as

$$H = \begin{pmatrix} \omega_r - i\gamma_\kappa - i\gamma_r & B & Je^{-i\beta L} \\ A - \kappa & \omega_r - i\kappa - i\gamma_r & Je^{i\beta L} \\ Je^{i\beta L} & Je^{-i\beta L} & \omega_{QD} - i\gamma_{QD} \end{pmatrix} \quad (1)$$

where $\omega_r$, $\omega_{QD}$ represent the mode frequency and the QD transition frequency; $\gamma_r$ is the intrinsic loss (due to leakage, absorption, or scattering), and $\gamma_{QD}$ denotes the decay into the free space continuum modes. $J$ is the constant value for optical fields at a particular transverse position along the propagation direction for a traveling-wave resonance; $\beta$ is the propagation constant, and $L$ is the relative distance between the QD and the reference point (e.g., the junction of the micro-ring and the WG). The Schematic of the QD-cavity system with the definitions of key physical parameters is illustrated in Fig. 1b. And the derivation of the Hamiltonian is detailed in SM section 2.

By solving the Schrodinger equation, the steady-state field amplitude distributions of the CW and CCW modes are obtained:

$$a_{EP} = \frac{Je^{-i\beta L}}{(\omega_{QD}-\omega_r)+i\gamma_\kappa+i\gamma_r} \quad (2)$$

$$b_{EP} = \frac{Je^{i\beta L}}{(\omega_{QD}-\omega_r)+i\gamma_\kappa+i\gamma_r} + \frac{-i2\gamma_\kappa|r|e^{i\varphi}Je^{-i\beta L}}{((\omega_{QD}-\omega_r)+i\gamma_\kappa+i\gamma_r)^2} \quad (3)$$

And the corresponding LDOS of the cavity is expressed as:

$$D_{EP}(\omega_{QD}) = -J^2[Im(a_{EP}(\omega_{QD})) + Im(b_{EP}(\omega_{QD}))] \quad (4)$$

As predicted, the coupling coefficient $\kappa = -2i\gamma_\kappa|r|e^{i\varphi}$ governs the formation of the EPs in this system. When |r| = 0, the system reduces to a conventional whispering-gallery-mode microcavity. Notably, the LDOS is modulated by the external phase term φ for |r| ≠ 0. Exotic quantum dynamics near EPs can be elucidated through the spectral properties of the LDOS, as is depicted in the right panel of Fig. 1c. When φ = 0, the LDOS line-shape of the EP cavity exhibits distinct splitting. The LDOS value is completely suppressed at the zero-detuning point (Δω = $\omega_r$ - $\omega_{QD}$= 0), which is termed EPIT (orange curve). For φ = 0.5π, the LDOS spectrum evolves into a Fano-like profile (green curve). When φ = π, the spectral shape transforms into a squared-Lorentzian profile (blue curve). The Lorentzian LDOS curve is plotted as a red dashed curve for comparison. The details about the theoretical modeling of EP-tailored LDOS profile are elaborated in the SM section 3.

The spontaneous emission spectrum of a QD in free space exhibits a prominent Lorentzian shape, with the linewidth determined by the lifetime of the QD. For a QD inside a cavity, its emission spectrum is modulated by the LDOS within the cavity. Compared to conventional optical microcavities, the rich LDOS emerging in EP cavities provides additional means to regulate single-photon emission. For our system, the coupling rate between the QD and the cavity is smaller than the decay rate of the cavity mode, which is usually referred to as the weak coupling regime. Under the Markov approximation, the spontaneous emission spectrum of the QD can be written as:

$$S_{\text{cavity}}(\omega) = J^2 \frac{1}{\pi} \frac{D_{EP}(\omega)(\gamma_{QD}/2)}{(\omega-\omega_{QD})^2+(\gamma_{QD}/2)^2} \qquad (5)$$

We calculate the modification of the QD's spontaneous emission spectrum inside the cavity as a function of the phase φ (see SM section 4). The results clearly show that the spontaneous emission rate and the related single-photon spectrum are modulated by the external phase φ. At zero detuning, when φ = 0, the single-photon emission completely vanishes because of null LDOS. For φ = 0.5π, the rate is the same as that of a DP cavity, except that the single-photon spectrum features a Fano-like profile instead of a Lorentzian line-shape. At φ = π, the LDOS is radically larger and the spontaneous emission rate achieves an enhancement factor twice that of the DP case (SM section 3). This makes the EP cavity particularly attractive for building high-performance on-chip quantum light sources, possibly enhancing their spontaneous emission rate beyond their current performance.

**Device design and fabrication**

The thin TFLN photonics has boosted applications from EO modulators to quantum optics (*38,39*). However, they are not optically active and incorporating solid-state quantum emitters in a TFLN platform remains technologically challenging. Solid-state platforms - particularly semiconductor QDs - offer unparalleled advantages through their giant dipole moments, small device footprint, and compatibility with industry nano-fabrication (*40*). Recent studies demonstrate cavity coupled QDs as high-performance single-photon sources (*41-44*) and ideal testbeds for exploring cQED, including strong light matter coupling (*45*) and collective radiation effects (*46*). Therefore, we adopt the hetero-integration method to take advantage of both materials. Additionally, strong Pockels electrooptic effect and piezoelectricity are proven to be preserved in TFLN even at low temperatures (*36,37*). This makes them particularly attractive in the photonic circuit integration of the QD single-photon sources which operate at the cryogenic temperature (*32*).

Fig. 2a shows the hetero-integrated quantum photonic chip. It consists of an x-cut LN resonator critically coupled to a LN waveguide with one end terminated by a Sagnac loop mirror. We employed the finite-difference time-domain (FDTD) method to numerically optimize the geometrical parameters of both LN and GaAs waveguides to ensure single mode operation (SM section 5). The cavity is a racetrack LN ring with a radius of 15 µm. The small ring size is designed to reduce the mode volume and thereby enhance the light-matter interaction. For the hetero-integration, we develop a method to utilize micro-sized polydimethylsiloxane (PDMS) stamps to reduce the printing error as described in SM section 6. A GaAs waveguide is precisely transfer-printed onto the LN resonator with a misalignment error of less than 100 nm as confirmed by the scanning electron microscopy image in Fig. 2b. To enable efficient evanescent coupling between the GaAs and LN waveguides, we implemented an adiabatic taper structure on the GaAs waveguide terminal as shown in Fig. 2c. Inset are the fundamental modes at the indicated cross sections. Considering practical nanofabrication constraints, the design adopts a taper tip width of 70 nm. Fig. 2d presents the calculated coupling efficiency as a function of taper length, revealing a rapid increase followed by saturation behavior. The efficiency reaches >90% at 10 µm taper length (dashed line). Based on these optimized parameters, the taper length was designed to 10 µm, ensuring high coupling efficiency while minimizing propagation losses (Inset of Fig. 2d).

To implement phase modulation (PM), two parallel metallic electrodes (length: 1 mm are patterned on either side of the LN waveguide, separated by a 5 µm gap to optimize the trade-off between EO tuning efficiency and optical propagation loss. The simulations indicate a half-wave voltage length product of $V\pi \cdot L = 1.4$ V · cm and a propagation loss of 0.1 dB/cm for this configuration (SM Section 7). Additionally, a pair of strain tuning electrodes is fabricated near the GaAs waveguide. Upon voltage application, the LN film undergoes mechanical deformation via the reverse piezoelectric effect, transmitting strain to the overlying GaAs waveguide. This strain selectively shifts the QD emission energy due to its strong bandgap sensitivity while leaving the cavity resonance frequency unaffected. To maximize strain tuning efficiency, the LN racetrack resonator is oriented at an 18° tilt relative to the crystallographic z-axis, to exploit the maximal piezoelectric coefficients in LN (*37*).

**Optical modes of the EP micro-cavity**

The chip is loaded in a closed-cycle cryostat operating at 5 K. Photoluminescence (PL) measurements were performed using a customized confocal microscopy system (see SM section 8). Notably, the grating coupler output exhibits much higher brightness compared to direct waveguide emission. confirming efficient inter-waveguide coupling. For the mode characterization, a high-power continuous-wave (CW) laser at 785 nm was focused onto the GaAs waveguide, with emitted signals collected through the grating coupler. Spectral analysis revealed a characteristic ring cavity resonance spectrum featuring equally spaced modes (free spectral range: $\Delta\lambda = 2$ nm consistent with the cavity design. The measured quality factor $Q = 1.4(1) \times 10^4$ is extracted from Lorentzian fits to cavity modes, demonstrating low optical losses in the hybrid photonic structure (Fig. 2e). Fig. 2f showcases the spectrum of a QD measured under a low excitation power, collected from the grating coupler. Inset shows the QDs' emission collected from GaAs waveguide top. The strong intensity contrast confirms that the radiation to the free space is negligible.

Application of voltage to the PM electrodes induces an EO tunable phase φ, which dynamically reconfigures the micro-ring cavity modes through controlled coupling between CW and CCW propagating waves. Fig. 2g presents the evolution of the mode spectrum upon varying phase φ. As discussed in the theoretical part, exotic quantum dynamics emerge. The mode evolves from EPIT to Fano-like to the squared-Lorentzian profile at phase parameters φ = 0, π/2, and π, respectively. The measured spectra are well reproduced by numerical calculations (Fig. 2h).

**On-chip local strain engineering**

Leveraging a precision transfer-printing technique, III-V GaAs waveguides incorporating QDs are monolithically integrated with LN photonics, achieving defect-free interfacial bonding via van der Waals forces. Achieving precise control of QD-cavity coupling in EP-engineered systems necessitates deterministic spectral alignment between QDs and microcavity resonances due to inherent fabrication variations in both QD emission wavelengths and micro-resonator geometries. This presents a critical challenge in integrated photonics. While existing approaches employ electrostatic actuation for local tuning - as demonstrated in diamond waveguide and nanocavity architectures - these methods suffer from limited tuning ranges (typically< 0.1nm) and complex multi-layer fabrication requirements (*33,34*).

To address these challenges, a monolithic microelectromechanical systems (MEMS) based strain tuning scheme is implemented, utilizing the reverse piezoelectric effect in LN. This functionality is enabled through lithographically defined electrodes adjacent to the hybrid III-V/LN waveguide on the X-cut LN chip (electrode geometry in inset of Fig. 3a). Given the anisotropy of strain coefficients in LN, the waveguide is tilted by 18° with respect to the crystallographic z-axis to optimize electromechanical coupling for enhanced tunability. Upon applying a bias voltage, the spectrum of single-photon emission is recorded. Resonant coupling at zero detuning manifests as a 15× enhancement in peak intensity (Fig. 3a). Fig. 3b is the spectrum mapping when the voltage is scanned from 0 V to 400 V. Experimental results demonstrate a linear wavelength tuning range of 0.6 nm across an applied voltage range, without affecting the cavity mode.

**Non-Hermitian single photon engineering**

Driven by the phase-dependent LDOS powered by the TFLN modulator, we experimentally demonstrate non-Hermitian cQED at the single-photon level. Spectra of the QD PL are recorded as the phase $\varphi$ is scanned from 0 to $4\pi$ (Fig. 4a). Fig. 4b shows the calculated QD's spontaneous emission spectrum as a function of the phase $\varphi$, which is in excellent agreement with experimental data. The measured data at $\varphi = 0$, $\varphi = 0.5\pi$ and $\varphi = \pi$ corresponding to the EPIT, Fano-like and squared Lorentzian spectrum are plotted in Fig. 4c. Insets are calculated curves under ideal conditions for comparison. The single-photon emission is regulated by the EP cavity, exhibiting similar features to the LDOS: the EPIT emerges when $\varphi = 0$. A transparent window occurs at the zero-detuning point. This is attributed to the finite QD emission linewidth ($\approx$0.03nm, comparable with the cavity linewidth $\approx$0.06nm). The emission couples to the LDOS shoulders around the transparent window (SM section 9). The spectrum evolves into a Fano-like profile at $\varphi = 0.5\pi$ and then transforms into a squared-Lorentzian profile for $\varphi = \pi$. The impacts of QDs' linewidth together with the non-unity mirror reflectivity and the intrinsic loss on the emission dynamics are discussed in SM section 10-11. The non-ideal conditions narrow down the EPIT window and make it eventually disappear with lower mirror reflectivity or larger intrinsic loss.

To elucidate the above observations, we further analyze the electric field distributions for key phase configurations via simulations. For $\varphi = 0$ (Fig. 4d), a perfect standing wave with a nodal point at the QD position effectively decouples the photonic mode from the emitter, suppressing the intensity to near-zero values. For $\varphi = 0.5\pi$ (Fig. 4e), the nodal point moves slightly away from the QD and couples the photonic mode to the emitter, resulting in a Purcell Factor the same as that of a DP cavity. Conversely, for $\varphi = \pi$ (Fig. 4f), an imperfect standing wave forms within the ring resonator, characterized by a CCW component amplitude threefold larger than the CW component. This field profile exhibits a maximum at the QD location, consistent with the observed intensity enhancement.

The single-photon characterization was performed using a Hanbury Brown-Twiss (HBT) interferometer comprising a 50:50 fiber beam splitter and single-photon detectors (SPDs). The measured second-order correlation function, $g^2(0) = 0.020(5)$ confirms high-purity single-photon emission, as illustrated in Fig. 4g. This result demonstrates near-ideal antibunching behavior, where $g^2(0) \ll 0.5$ validates the suppression of multi-photon events - a critical requirement for quantum information applications.

We then performed time-resolved PL measurements to reveal the spontaneous emission dynamics. The lifetime of QD is shortened from 850 ps to 120 ps when φ is tuned from 0 to π. The corresponding decay curves together with deconvolution fittings are plotted in Fig. 4h. These changes can be explained by the LDOS, which minimizes at EPIT and maximizes in the squared-Lorentzian regime. The radiation rate at various φ is summarized in Fig. 4I, which shows a sinusoidal dependence on φ. Notably, even within the EPIT regime, the QD emission lifetime ($\tau$ = 850ps) remains significantly shorter. This phenomenon again originates from the spectral coupling between single-photon transitions and the LDOS shoulders in the EPIT regime (SM section 9). Strict zero-linewidth conditions would lead to complete suppression of spontaneous emission, as dictated by the vanishing density of optical states at spectral nodes.

**Conclusion**

By leveraging LN's dual electro-optic and piezoelectric properties, we achieve simultaneous control over feedback phase φ and QD emission wavelength, enabling real-time modulation of Purcell factors and LDOS profiles. A 7-fold dynamic range in the lifetime (850 ps to 120 ps) is demonstrated, with potential GHz-scale switching speeds through traveling-wave electrode designs (*47*). Such capabilities could enable ultrafast quantum gates (*48,49*) or single-photon routers in integrated circuits (*50*), addressing critical bottlenecks in photonic quantum computing. The demonstrated work bridges non-Hermitian photonics and cQED by establishing the first experimental platform for EP-engineered single-photon dynamics. The phase-dependent LDOS profiles allow spectral shaping of single photons into non-classical waveforms - a capability absent in Hermitian systems. The observed phenomena: squared-Lorentzian line-shapes, Fano-like spectrum, and EPIT directly originate from the singular topology of EPs. This mechanism unlocks a new regime of single-photon control, where emission can be suppressed or enhanced purely through cavity topology rather than mode matching, and offers exciting opportunities to explore all-electrical reconfigurable quantum light sources for advanced quantum photonic science and technology.

**References**

1 H.-Z. Chen, T. Liu, H.-Y. Luan, et al. Revealing the missing dimension at an exceptional point. *Nature Physics* **16** (5), 571–578, (2020)
2 C. Wang, X. Jiang, G. Zhao, et al. Electromagnetically induced transparency at a chiral exceptional point. *Nature Physics* **16** (3), 334–340, (2020)
3 B. Peng, S¸. K. Ozdemir , S. Rotter, et al. Loss-induced suppression and revival of lasing *Science* **346**, 328–332, (2014)
4 A. Li, H. Wei, M. Cotrufo, et al. Exceptional points and nonHermitian photonics at the nanoscale. *Nature Nanotechnology* **18** (7), 706–720, (2023)
5 H. Du, X. Zhang, C. G. Littlejohns, et al. Nonconservative Coupling in a Passive Silicon Microring Resonator. *Phys. Rev. Lett.* **124** (1), 013606, (2020)
6 S. Soleymani, Q. Zhong, M. Mokim, et al. Chiral and degenerate perfect absorption on exceptional surfaces. *Nature Communications* **13** (1), 599, (2022)
7 Z. Shen, Y.-L. Zhang, Y. Chen, et al. Experimental realization of optomechanically induced non-reciprocity. *Nature Photonics* **10** (10), 657–661, (2016)
8 Q. Zhong, J. Ren, M. Khajavikhan, et al. Sensing with Exceptional Surfaces in Order to Combine Sensitivity with Robustness. *Phys. Rev. Lett.* **122** (15), 153902, (2019) 9 Kejie Fang, Zongfu Yu and Shanhui Fan. Photonic AharonovBohm Effect Based on Dynamic Modulation. *Phys. Rev. Lett.* **108**, 153901, (2012)

10 Cohen, E. et al. Geometric phase from Aharonov–Bohm to Pancharatnam–Berry and beyond. *Nat. Rev. Phys.* **1**, 437–449 (2019)

11 Enbang Li, Benjamin J. Eggleton, Kejie Fang et al. Photonic Aharonov–Bohm effect in photon–phonon interactions. *Nat. Commun.* **5**,3225 (2014)

12 B. Peng, S. K. Ozdemir, M. Liertzer, et al. Chiral modes and directional lasing at exceptional points. *Proc. Natl. Acad. Sci.* **113** (25), 6845–6850, (2016)

13 Hwaseob Lee, Lorry Chang, Ali Kecebas et al. Chiral exceptional point enhanced active tuning and nonreciprocity in microresonators. *light: science & applications* **14**,45 (2025)

14 Y. Chen, J. Li, K. Xu et al. Electrically reconfigurable mode chirality in integrated microring resonators. *Laser & Photonics Reviews* **18**,2301289 (2024)

15 William R. Sweeney, Chia Wei Hsu, Stefan Rotter, et al. Perfectly Absorbing Exceptional Points and Chiral Absorbers. *Phys. Rev. Lett.* **122**, 093901, (2019)

16 C. Wang, W. R. Sweeney, A. D. Stone, et al. Coherent perfect absorption at an exceptional point. *Science* **373** (6560), 1261– 1265, (2021)

17 Shai Maayani, Raphael Dahan, Yuri Kligerman et al. Flying couplers above spinning resonators generate irreversible refraction. *Nature*, **558**, 569-572 (2018)

18 J. Xu, Y. Mao, Z. Li, et al. Single-cavity loss-enabled nanometrology. *Nature Nanotechnology* **19** (10), 1472–1477, (2024)

19 Y.-P. Ruan, J.-S. Tang, Z. Li, et al. Observation of loss-enhanced magneto-optical effect. *Nature Photonics* **19** (1), 109–115, (2025)

20 W. Mao, Z. Li, Y. Li, et al. Exceptional-point-enhanced phase sensing. *Science Advances* **10** (14), ead15037, (2024)

21 Z. Zhang, X. Qiao, B. Midya, et al. Tunable topological charge vortex microlaser. *Science* **368** (6492), 760–763, (2020)

22 G. Madiot, Q. Chateiller, A. Bazin, et al. Harnessing coupled nanolasers near exceptional points for directional emission. *Science Advances* **10** (45), eadr8283, (2024)

23 A. K. Jahromi, A. U. Hassan, D. N. Christodoulides, et al. Statistical parity-time-symmetric lasing in an optical fibre network. *Nature Communications* **8** (1), 1359, (2017)

24 L. Ferrier, P. Bouteyre, A. Pick, et al. Unveiling the Enhancement of Spontaneous Emission at Exceptional Points. *Phys. Rev. Lett.* **129** (8), 083602, (2022)

25 Ran Huang, Adam Miranowicz, Jie-Qiao Liao et al. Nonreciprocal Photon Blockade. *Phys. Rev. Lett.* **121**, 153601 (2023)

26 Xingda Lu, Wanxia Cao, Wei Yi et al. Nonreciprocity and Quantum Correlations of Light Transport in Hot Atoms via Reservoir Engineering. *Phys. Rev. Lett.* **126**, 223603 (2021)

27 Zimo Zhang, Zhongxiao Xu, Ran Huang et al. Chirality-induced quantum nonreciprocity *Arxiv*: 2504.13437 (2025)

28 Q. Zhong, A. Hashemi, S. K. Ozdemir, et al. Control of spontaneous emission dynamics in microcavities with chiral exceptional surfaces. *Phys. Rev. Research* **3** (1), 013220, (2021)

29 A. Pick, B. Zhen, O. D. Miller, et al. General theory of spontaneous emission near exceptional points. *Opt. Express* **25** (11), 12325–12348, (2017)

30 M. Khanbekyan, J. Wiersig. Decay suppression of spontaneous emission of a single emitter in a high-Q cavity at exceptional points. *Phys. Rev. Research* **2** (2), 023375, (2020)

31 M. Davanco, J. Liu, L. Sapienza, et al. Heterogeneous integration for on-chip quantum photonic circuits with single quantum dot devices. *Nature Communications* **8** (1), 889, (2017)

32 S. Aghaeimeibodi, B. Desiatov, J.-H. Kim, et al. Integration of quantum dots with lithium niobate photonics. *Appl. Phys. Lett.* **113** (22), 221102, (2018) 33 N. H. Wan, T.-J. Lu, K. C. Chen, et al. Large-scale integration of artificial atoms in hybrid photonic circuits. *Nature* **583** (7815), 226–231, (2020)
34 H. Larocque, M. A. Buyukkaya, C. Errando-Herranz, et al. Tunable quantum emitters on large-scale foundry silicon photonics. *Nature Communications* **15** (1), 5781, (2024)
35 A. W. Elshaari, W. Pernice, K. Srinivasan, et al. Hybrid integrated quantum photonic circuits. *Nature Photonics* **14** (5), 285–298, (2020)
36 E. Lomonte, M. A. Wolff, F. Beutel, et al. Single-photon detection and cryogenic reconfigurability in lithium niobate nanophotonic circuits. *Nature Communications* **12** (1), 6847, (2021)
37 W. Yue, J. Yi-jian. Crystal orientation dependence of piezoelectric properties in LiNbO3 and LiTaO3. *Optical Materials* **23** (1), 403–408, (2003)
38 A. Boes, L. Chang, C. Langrock, et al. Lithium niobate photonics: Unlocking the electromagnetic spectrum. *Science* **379** (6627), eabj4396, (2023)
39 C. Wang, M. Zhang, X. Chen, et al. Integrated lithium niobate electro-optic modulators operating at CMOS-compatible voltages. *Nature* **562** (7725), 101–104, (2018)
40 R. Uppu, L. Midolo, X. Zhou, et al. Quantum-dot-based deterministic photon-emitter interfaces for scalable photonic quantum technology. *Nature Nanotechnology* **16** (12), 1308–1317, (2021)
41 X. Ding, Y.-P. Guo, M.-C. Xu, et al. High-efficiency singlephoton source above the loss-tolerant threshold for efficient linear optical quantum computing. *Nature Photonics* **19**, 387–391, (2025)
42 N. Tomm, A. Javadi, N. O. Antoniadis, et al. A bright and fast source of coherent single photons. *Nature Nanotechnology* **16** (4), 399–403, (2021)
43 N. Somaschi, V. Giesz, L. De Santis, et al. Near-optimal singlephoton sources in the solid state. *Nature Photonics* **10** (5), 340–345, (2016)
44 R. Uppu, F. T. Pedersen, Y. Wang, et al. Scalable integrated single-photon source. *Science Advances* **6** (50), eabc8268, (2020)
45 Daniel Najer, Immo Sollner, Pavel Sekatski et al. A gated quantum dot strongly coupled to an optical microcavity. *Nature* **575**, 622-627 (2019)
46 A.Tiranov, V. Angelopoulou, et al. Collective super- and subradiant dynamics between distant optical quantum emitters. *Science* **379**, 389-393 (2023)
47 C. Wang, M. Zhang, X. Chen et al. Integrated lithium niobate electro-optic modulators operating at CMOS-compatible voltages *Nature* **562**, 101–104, (2018)
48 C.-Y. Jin, R. Johne, M. Y. Swinkels, et al. Ultrafast non-local control of spontaneous emission. *Nature Nanotechnology* **9** (11), 886–890, (2014)
49 Gary Shambat, Bryan Ellis, Arka Majumdar et al. Ultrafast direct modulation of a single-mode photonic crystal nanocavity lightemitting diode. *Nat. Commun.* **2**, 539 (2011)
50 Camille Papon, Xiaoyan Zhou, Henri Thyrrestrup et al. Nanomechanical single-photon routing. *Optica* **6**, 524-530 (2019)

**Acknowledgments:** We thank Prof. Hui Jing for the fruitful discussion.

**Funding:** This research is supported by the following grants:

National Natural Science Foundation of China 12374476 (Y.C.)

National Natural Science Foundation of China 62422503,12474375 (J.W.)

National Natural Science Foundation of China 12474369 (R.S.)

The Chinese Academy of Sciences Project for Young Scientists in Basic Research YSBR-112 (J.Z.)

The Strategic Priority Research Program of the Chinese Academy of Sciences XDB0670303 (J.Z.)

Guangdong Introducing Innovative and Entrepreneurial Teams of "The Pearl River Talent Recruitment Program" 2021ZT09X044 (J.L.)

The innovative research program 22-ZZCX-067 (Y.C.)

Guangdong Basic and Applied Basic Research Foundation 2023B1515120070 (J.L.)

**Author contributions:**

Conceptualization: YC, JW, JT, JL

Methodology: YC, JZ, RS

Investigation: XW, KX, XL, KW, TZ, YY, XL, YC

Visualization: YC, JL, JW

Funding acquisition: YC, TJ

Supervision: TJ, JL

Writing – original draft: YC, JW

Writing – review & editing: JL

**Competing interests:** The authors declare that they have no competing interests.

**Data and materials availability:** All data are available in the main text or the supplementary materials.

## Supplementary Materials

Supplementary Text

Figs. S1 to S12

References (Ref. S1-S6)

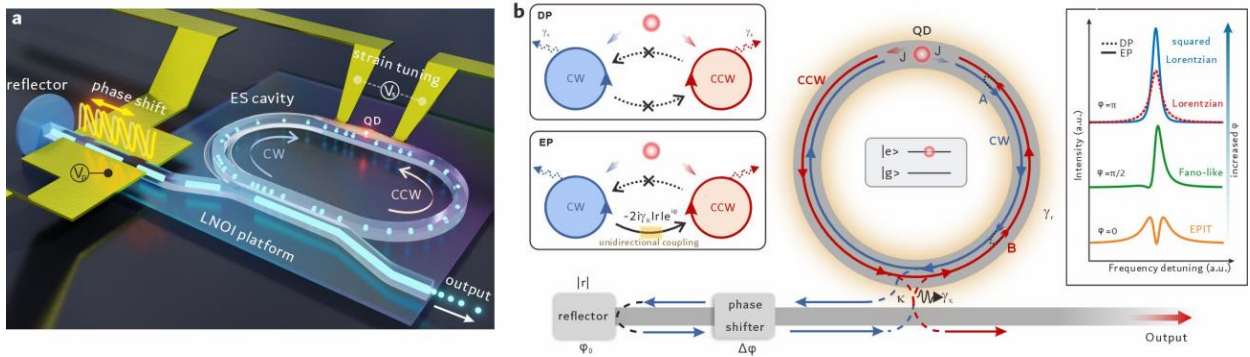

**Fig. 1. A dynamically tunable quantum photonic platform based on a chiral EP microcavity.** (**a**) Artistic sketch of the proposed device. The device comprises a LN micro-ring resonator evanescently coupled to a waveguide terminated by a mirror. A GaAs waveguide embedded with a strain-tunable QD - a two-level quantum emitter - is transfer-printed onto the LN resonator, forming a hybrid integrated circuit. The QD's emission wavelength is strain-tuned via the piezoelectric effect to align with the cavity resonance. EO phase modulation dynamically adjusts the coupling from CW to counterclockwise CCW cavity modes. (**b**) Schematic of the QD-cavity system with the definitions of key physical parameters. The QD couples to a DP cavity, where the CW mode and CCW mode are degenerated and have no mutual coupling. While in an EP cavity, the CCW mode unidirectionally couples to the CW mode (upper left). The DP cavity is a Lorentzian curve. And the EP exhibits exotic spectral responses including EPIT, Fano-like and squared-Lorentzian line-shape under phase modulation (right panel).

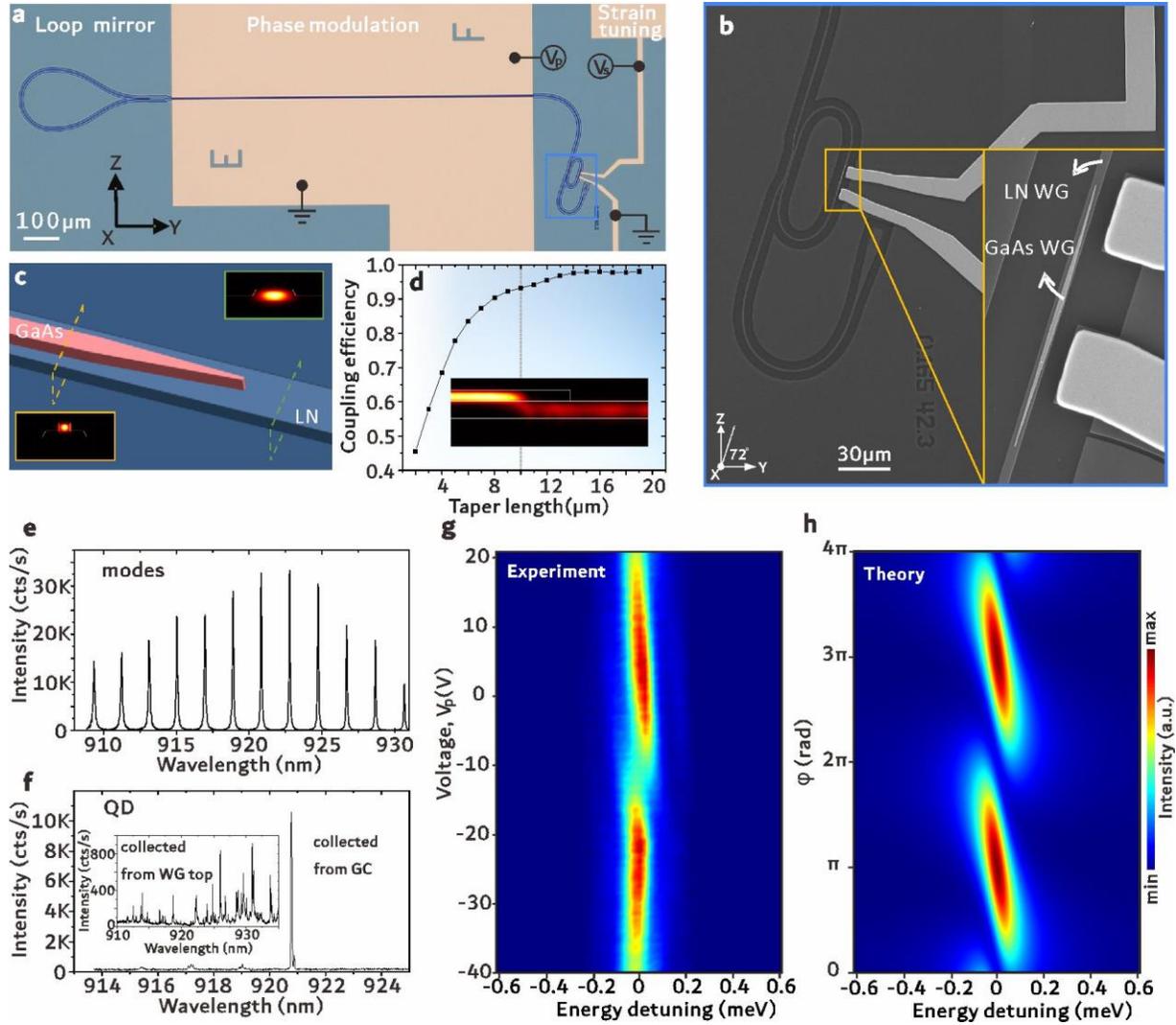

**Fig. 2. Hybrid quantum photonic chip and chiral EP cavity mode engineering.** (**a**) Fabricated hybrid LN-GaAs photonic chip, comprising three functional sections: loop mirror, electro-optic phase modulation, and the strain tuning. (**b**) Zoomed-in of the cavity region: A tapered GaAs waveguide with QDs is transfer-printed onto the LN resonator. The scanning electron microscopy (SEM) image confirms high transfer accuracy (Inset). (**c**) The sketch of the hetero-integration scheme and mode profiles at indicated sections. (**d**) The coupling efficiency as a function of taper length, which reaches above 90% for 10µm. Inset is the simulated light propagation from the GaAs waveguide to the LN waveguide. (**e**) Spectrum of cavity modes. (**f**) Spectrum of QDs excited with a low laser power. The spectrum is collected from the grating coupler (GC). Inset shows the QDs' emission collected from GaAs waveguide top. (**g**) Measured and (**h**) theoretical mode spectra evolution under phase modulation ($\varphi = 0$ to $4\pi$), demonstrating transitions between Lorentzian, Fano-like, and exceptional-point induced transparency (EPIT) regimes.

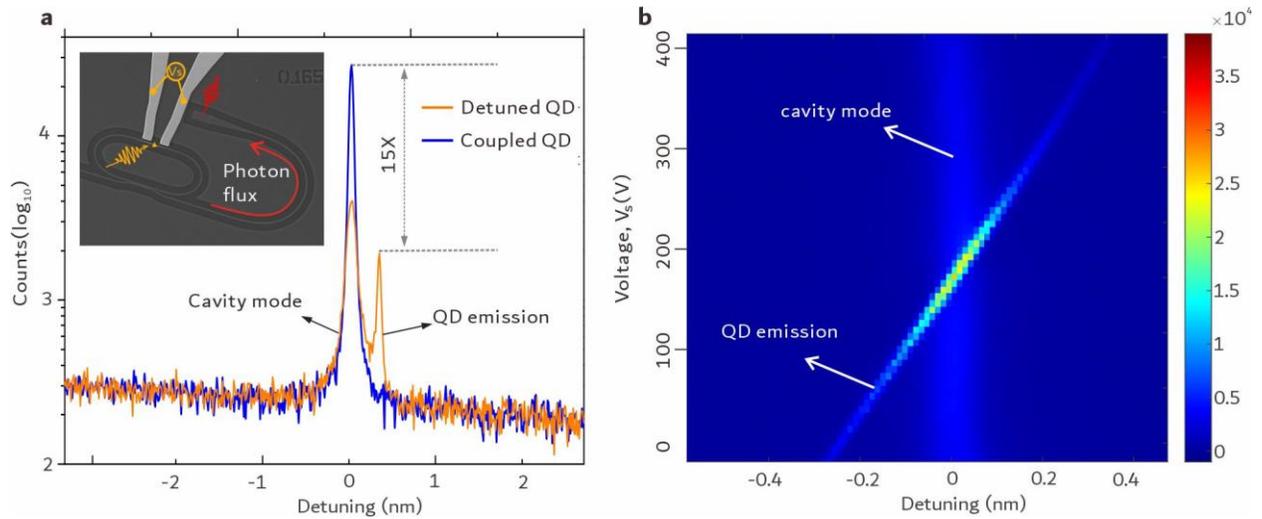

**Fig. 3. On-chip local strain-tuning of QD emission.** (**a**) Spectra of a QD under detuned and resonant s. A 15-fold enhancement of the emission intensity is observed when tuning the QD into the cavity mode. Voltage is applied to the two electrodes adjacent to the GaAs waveguide (Inset). (**b**) QD emission is spectrally scanned through the cavity resonance using the reverse piezoelectric effect (0–400 V). A total wavelength shift of 0.6 nm is observed.

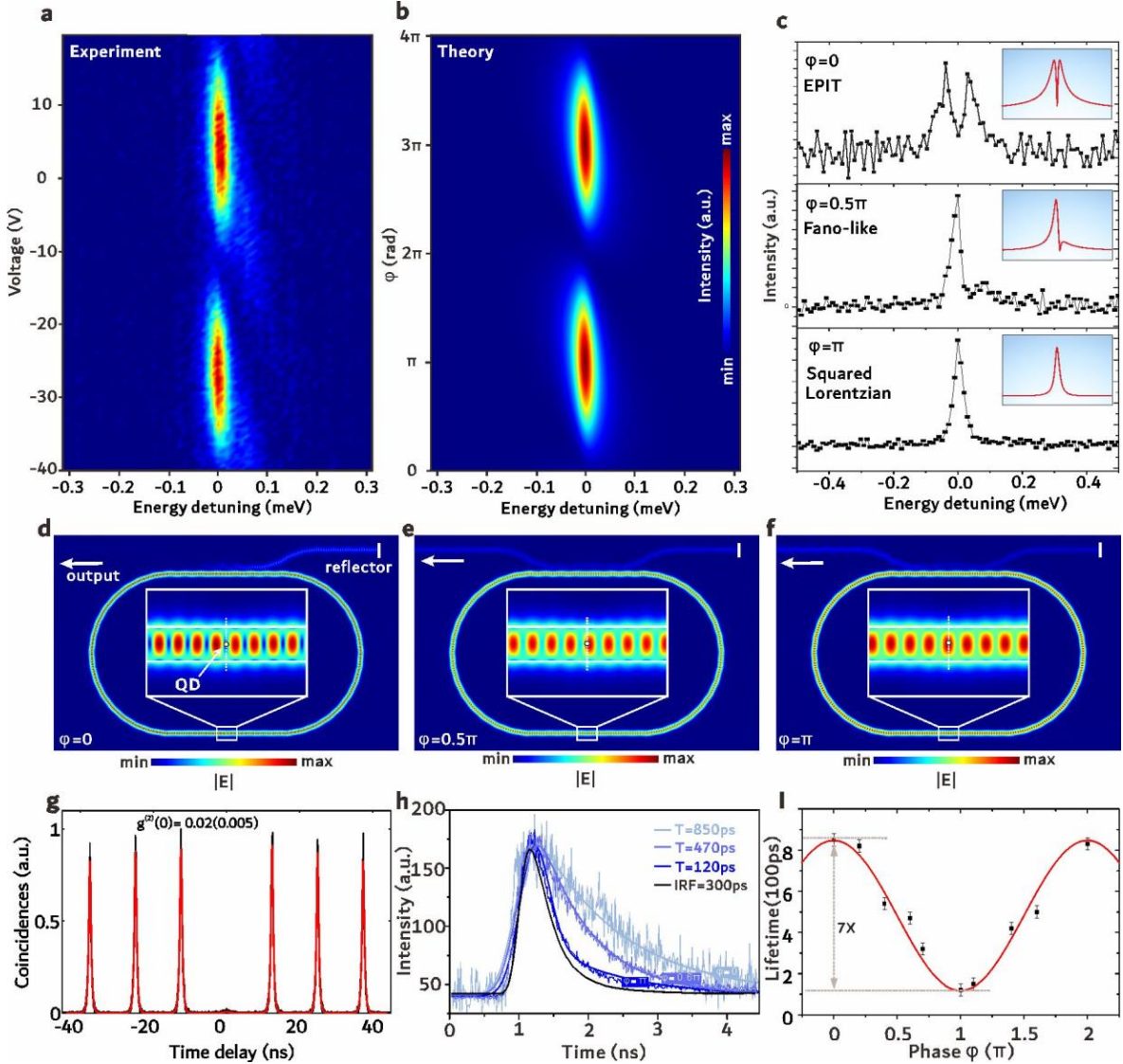

**Fig. 4. Non-Hermitian control of single-photon emission dynamics via phase modulation.** (**a**) Experimentally measured and (**b**) calculated single-photon spectra under continuous phase modulation ($\varphi = 0$ to $4\pi$), (**c**) The QD spectral slices at $\varphi = 0$ (EPIT), $\pi/2$ (Fano-like), and (squared-Lorentzian). Insets are calculated curves under ideal conditions. (**d-f**) Distribution of the electric field in the structure under semi-steady state conditions for the three scenarios. Insets show the QD's position with respect to the field distribution. (**d**) $\varphi = 0$. The CW and CCW waves form a perfect standing wave pattern with the null located at the position of the QD. (**e**) $\varphi = \pi/2$. The QD sits at the edge of the node of the electric field. The Purcell Factor (PF) enhancement is the same as that of a DP cavity. (**f**) $\varphi = \pi$. The PF enhancement maximizes because of larger field amplitude; (**g**) Second-order correlation measurement using a Hanbury Brown-Twiss (HBT) interferometer. The suppressed peak at zero-time delay ($g2(0) = 0.02(0.005)$) confirms high single-photon purity, with multiphoton probability of 0.02. (**h**)Time resolved measurement: The lifetime traces of slow and fast decay curves together with deconvolution fittings, are plotted. The instrument response function (IRF) is represented by the black line. (**I**) The lifetime as a function of phase modulation. Lifetime tuning spans from 120 ps to 850 ps, achieving a 7-fold dynamic range.